\documentstyle[lprocl,11pt]{article}
\input{psfig}
\def\Journal#1#2#3#4{{#1} {\bf #2}, #3 (#4)}

\def\PRL{\em Phys. Rev. Lett.}
\def\PRL{\em Phys. Rev. Lett.}
\def\PR{\em Phys. Rev.}

\def\ZPB{{\em Z. Phys.} B}
\def\JPA{{\em J. Phys.} A}
\def\JPC{{\em J. Phys.} C}

\def\CMP{\em Comm. Math. Phys.}
\def\JSP{\em J. Stat. Phys.}
\def\EL{\em Europhys. Lett.}
\def\JP{\em J. Physique}
\def\JdPI{{\em J. de Physique} I}
\def\JdPII{{\em J. de Physique} II}
\def\PT{\em Phase Transition}
\def\PRB{{\em Phys. Rev} B}

\def\PA{{\em Physica} A}
\def\RMP{\em Rev. Mod. Phys.}
\def\IJMP{\em Int. J. Mod. Phys.}
\def\JMMM{\em J. Magn. Mater}
\def\JPL{{\em J. Physique Lett} (France)} 

\def\mco{\multicolumn}

\def\be{\begin{equation}}
\def\ee{\end{equation}}
\def\bea{\begin{eqnarray}}
\def\eea{\end{eqnarray}}

\begin{document}
\title{THEORY OF THE RANDOM FIELD ISING MODEL}
\author{ T. NATTERMANN}
\address{Institute for Theoretical Physics, University of Cologne,
Z\"ulpicherstr. 77, 50937 Cologne, Germany}
\maketitle\abstracts{A review is given on some recent developments 
in the theory of the 
Ising model in a random field. This model is a good representation 
of a large number of impure materials. After a short repetition of 
earlier arguments, which prove the absence 
of ferromagnetic order in $d\le 2$ space dimensions for 
uncorrelated random fields, 
we consider different random field 
correlations and in particular the generation of uncorrelated 
from anti-correlated random fields by thermal fluctuations. 
In discussing the 
phase transition, we consider the transition to be characterized by 
a divergent correlation length and compare 
the critical exponents obtained from various methods (real space RNG, 
Monte Carlo calculations, weighted mean field theory etc.). The 
ferromagnetic transition is believed to be preceded by a spin glass 
transition which manifests itself by replica symmetry breaking. 
In the discussion of dynamical properties, we concentrate mainly on 
the zero temperature depinning transition of a domain wall, which 
represents a critical point far from equilibrium with new scaling relations 
and critical exponents. }

\section{Introduction}

The Lenz--Ising model is probably the oldest and most simple non--trivial
model for cooperative behavior which shows spontaneous symmetry 
breaking~\cite{Ising,Domb}.
It has a vast number of applications ranging from solid state physics to
biology. Its Hamiltonian is given by
\begin{equation}
{\cal H}=-\sum\limits_{i,j}J_{ij}S_iS_j-\sum\limits_iH_iS_i
\label{eq:Ham}
\end{equation}
where $S_i=\pm 1$. In a more restricted sense, the Ising model is
understood to have coupling constants which are translationally invariant, 
$J_{ij}=J(|{\bf r}_i-{\bf r}_j|)$, where ${\bf r}_i,\, {\bf r}_j$
denote points of a regular $d$--dimensional lattice. The external field 
$H_i$ is typically homogeneous, $H_i\equiv H({\bf r}_i)=H\, \forall i$, 
or depends only
smoothly on ${\bf r}_i$. Even under this restricted conditions the Ising
model exhibits a remarkable complexity. In particular, if competing
interactions are taken into account, the Ising model may show a large 
number of commensurate and incommensurate phases of modulated 
magnetization~\cite{Selke}.
In reality, systems are never completely translationally invariant.
Compositional disorder, impurities and vacancies, lattice dislocations etc.
lead to modifications in the Hamiltonian, which in many cases may
be characterized by changes 
$J(|{\bf r}_i-{\bf r}_j|) \rightarrow J(|{\bf r}_i-{\bf r}_j|)+ \delta J_{ij}$
and $H \rightarrow H+h_i$ in ${\cal H}$. $\delta J_{ij}$ and $h_i$ 
are no longer translationally invariant,
but random quantities, characterized by their probability
distributions. Typically the values of $\delta J_{ij}$ and $h_i$
have a zero average and
are uncorrelated for different bonds and sites,
respectively. Let us briefly consider the different limiting cases
separately.

For $h_i\equiv 0$ and $\delta J_{ij}\ll J_{ij}$, we expect, that the
influence of the randomness is not very dramatic. Higher order
commensurate phases may disappear~\cite{Bak}. In the case of second order
transitions the critical exponents can be changed~\cite{Harris}.
First order transitions on the other hand may become second order~\cite{Berker}.

If $\delta J_{ij}\gg J_{ij}$ however, ferromagnetic order will in
general be destroyed. A low temperature spin glass phase with 
$\overline{<S_i>}=0$ but $\overline{<S_i>^2}=q_{EA}>0$ may occur~\cite{BiYou}.
Here  $<\ldots >$ denotes the thermodynamic average and the overbar
the average over the disorder configurations. The latter replaces the
spatial average over the sample.

In this article we will consider the complementary 
case $\delta J_{ij}\equiv 0$ and $h_i\not= 0$, i.e. the Ising model in a random
field.
If not stated otherwise we will assume below
\begin{equation}
\overline{h_i}=0\quad,\quad\overline{h_ih_j}=h^2\delta_{ij}\quad\hbox{and}\quad h\ll J.
\label{eq:hi,hihj} 
\end{equation}
The exchange constant $J_{ij}=J$ is assumed to be short ranged and non-zero 
between nearest neigbors $<i,j>$ only.
An alternative soft--spin description of the random field
Ising model is given by the Ginzburg--Landau--Hamiltonian
\begin{equation}
{\cal H}_{\rm GL}=\int {\rm d}^dr\,\left\{\frac{1}{2}{\tilde a}\phi^2+\frac{1}{2}
(\nabla \phi )^2+\frac{1}{4}u\phi^4 -h({\bf r})\phi ({\bf r})\right\}
\label{eq:HGL}
\end{equation}
with $ -\infty < \phi({\bf r}) < \infty , {\tilde a}=a'(T-T_c(0)), 
T_c(h=0)=O(J), u>0$ and
\begin{equation}
\overline{h({\bf r})}=0\,,\quad \overline{h({\bf r})h({\bf r}')}=
h^2\delta ({\bf r}-{\bf r}') .
\label{eq:barhr}
\end{equation}

Since its seminal discussion by Imry and Ma~\cite{IM} in 1975,
this model is under intensive investigation both experimentally
and theoretically. Results obtained before 1991 are summarized 
to a some extend in earlier reviews~\cite{reviews1,reviews3,reviews2,Rieger}. 
In the present paper we will therefore mainly concentrate 
on more recent findings and refer to the earlier work only whenever needed
to make the paper more self-contained.

As has been discussed earlier, the random field Ising model has a number
of interesting realizations in nature. The most studied experimental
systems are diluted antiferromagnets in a homogeneous external 
field where  the combination of dilution and external field 
leads to a random field like effect for the staggered 
magnetization~\cite{FishAha}.

Another example is adsorbed mono-layers on impure substrates, e.g. $Xe$ on a 
$Cu(110)$ surface. Here the mono-layer has two (or more) ground-states on the 
substrate. If one of the substrate lattice sites is occupied by a frozen--in 
impurity,
 it prevents additional occupation of this site with an 
ad--atom. Thus it acts locally as a symmetry breaking field~\cite{Vill82}.

Further realizations of random field systems are binary liquids in 
porous media~\cite{deGennes}, mixed Jahn--Teller systems~\cite{Graham},
diluted frustrated antiferromagnets~\cite{Fernand}, hydrogen
in metals~\cite{Peisl} and mixed crystals undergoing
structural or ferroelectric transitions~\cite{Michel}. 
Recently, an application of the random field Ising model for the 
understanding of cooperativity of protein folding has been
discussed by Shakhnovich and coworkers~\cite{Gut} . Another more recent 
development is the discussion of  the 
Anderson-Mott transition of disordered interacting electrons 
as a random field problem~\cite{KirkBelitz}.

The rest of the paper is organized as follows: In Section 2 we consider the 
stability  of the ferromagnetically ordered phase, in particular in the 
presence of field correlations which deviate from Eq. \ref{eq:hi,hihj}. 
Section 3 is devoted to the discussion of the critical behavior. In Section 4 
we discuss some dynamical properties of the model. 
Finally, Section 5 is reserved for 
miscellaneous subjects related to the random field problem.

\section{Ordered Phases}
\subsection{The stability of the ferromagnetic phase}\label{subsec:ferromphase}

The pure Ising model $ {\cal H}_0=-J\left.\sum\right._{<ij>}S_iS_j$
is known to have a ferromagnetically ordered phase in all dimensions $d > 1$.
An additional random field term ${\cal H}_1=-\left.\sum\right._ih_iS_i$
will act against this order. When the random field strength $h$ is
sufficiently large compared to $J$, the system is disordered at low 
temperatures, as shown in an exact treatment~\cite{BE,Froe}. The opposite
case $h\ll J$ is of more interest. As it was first shown by Imry and 
Ma~\cite{IM},
the ferromagnetic ground-state becomes indeed unstable with respect
to the formation of ill--oriented domains in all dimensions $d\le 2$.

\begin{figure}[h]
\hspace{5.5cm}
\psfig{figure=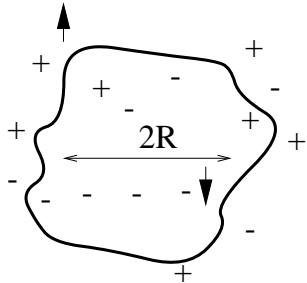,height=1.5in}
\caption{domain of reversed spins}
\end{figure}

If one reverses the spins (Fig. 1) within a domain of size $R$,
the energy cost $E_{\rm ex}$ is proportional to the domain
wall area
$E_{\rm ex}\sim J R^{d-1}$. For simplicity, we set the lattice constant 
equal to unity.
This energy increase has to be compared with the energy gain from the 
interaction
with the random field. Clearly, the average Zeeman energy vanishes for the 
ferromagnetic state. However, according to the central  limit theorem,
the mean square random field energy $E^2_{\rm RF}$ inside a region
of volume $R^d$ is of the order $h^2 R^d$. The energy $E_{\rm RF}$
for a given region may be positive or negative with equal
probability. It is always possible however to find a region
enclosing an arbitrary point $i$ such that $E_{\rm RF}>0$.
Reversing the spins in this region yields an energy gain of
$-2E_{\rm RF}$. The total energy of the domain of size $R$ is
therefore
\begin{equation}
E(R)\approx J R^{d-1}-h R^{d/2} .
\label{eq:ER}
\end{equation}
For $h\ll J$, $E(R)$ is positive for $d\ge 2$, but negative for $d<2$
if $R$ is large enough. Thus the ferromagnetic state becomes
unstable with respect to domain formation for $d<2$. Later, Binder~\cite{Bi}
was able to show, that in $d=2$ dimensions the roughness of the domain
surface generated by the random fields leads to an instability
of the surface tension for 
$R\approx R_c \sim {\exp{(J/h)}}^2$ and hence 
there is no long range order also in $d=2$ dimensions. 
Aizenman and Wehr~\cite{AizW} proved indeed rigorouly, that in $d\le 2$ 
dimension the random field produces a unique Gibbs state, i.e. 
the absence of any phase transition, in agreement with the naive expectation 
from the Imry-Ma argument. In higher dimensions 
the surface roughness has no influence on the long range order. 

So far we have assumed, that domains are compact and do not include 
smaller domains of reversed spins. Moreover, entropic effects were 
neglected. Indeed, it could be shown, that 

(i) domains within domains, which would renormalize $h\rightarrow h(R)$,
are rare if $d>2$ and can hence be neglected~\cite{Imb},

(ii) although there is a large number $\tilde{N}(R)\sim\exp{(c R^{d-1})}$ of
contours for a domain of size $R$, these enclose essentially
the same volume and the same random fields. Here $ c$ is an unknown 
numerical coefficient. Thus, with probability one there 
is no contour which encloses a random field gain which is larger than
the surface energy loss if $d>2$~\cite{FFS},

(iii) thermal fluctuations are irrelevant at low temperatures and can hence
be neglected~\cite{BC}.

Thus,
one concludes, that the lower critical dimensions for the ferromagnetically
ordered phase is $d_l=2$. 

\subsection{Different field correlations}\label{subsec:lri}

So far we have assumed, that both the interaction between spins and the 
correlations between the random fields are short range. In this
paragraph we will briefly consider the extension of the Imry--Ma
argument to long range interaction and correlations, respectively.

Let us first consider the case of {\em infinite--range interaction},
${\cal H}_0=-(J/N)\left.\sum\right._{i<j}S_iS_j$. Here $N$ denotes the 
total number of spins in the system. The energy increase by
flipping a domain is now of the order $J R^d$ and hence always
larger than the energy gain from the random field. The system is
therefore ordered at low temperatures in all dimensions.

Another type of long range interaction comes from {\em dipolar forces}. For
nearly spherical domains these lead to an additional term $g R^{2d-3}$ in 
Eq. \ref{eq:ER},
where we assumed three dimensional dipolar interaction
(vanishing as $R^{-3}$) between dipoles arranged in a $d$--dimensional 
space~\cite{Natter88}. $g$ is a measure of the strength of the dipolar 
interaction.
It is easy to show, that this term does not change the lower critical
dimension $d_l=2$. In general, domains may not be spherical and
indeed will tend to take a cigar--like shape with the long axis
parallel to the spin direction in order to lower demagnetisation
factors, but this will not change $d_l$ either.
A slightly different treatment is necessary for diluted antiferromagnets
in a homogeneous external field, which are a good experimental realization
of the random field Ising model, but the conclusion $d_l=2$
applies also to this case~\cite{Natter88}.

\medskip

The situation becomes different if we allow {\em long--range correlations}
of the random fields, i.e.
\begin{equation}
\overline{h_ih_j} =h^2|{\bf r}_i-{\bf r}_j|^{d'-d}\quad\hbox{for $i\neq j$ and $\overline{h_i^2}=h^2$}\,.
\label{eq:hihj1}
\end{equation}
For $d'>0$ the lower critical dimension changes to $d_l=2+d'$~\cite{Natter83}. Examples
are random fields correlated along lines $(d'=1)$ or planes $(d'=2)$. 
A special example is the $d$--dimensional quantum 
ferroelectric with uncorrelated random fields at $T=0$, which
can be described by a $(d+1)=D$--dimensional classical model and
$d'=1$, since the disorder is correlated in the (imaginary) time
direction. Thus $D_l=2+1=3$ and hence $d_l=D_l-1=2$,
i.e. quantum effects do not change $d_l$ in this case~\cite{AGS}.

An opposite situation exists in systems with {\em anti-correlated
fields} \cite{SchVShN}
\begin{equation}
\overline{h_ih_j} =\left\{ \begin{array}{c@{\quad,\quad}l}
-c(\mu )h^2|{\bf r}_i-{\bf r}_j|^{-(d+2\mu )} & i\not= j\\
h^2 & i=j
\end{array}\right.
\label{eq:hihj2}
\end{equation}
where $c(\mu )$ is a constant determined by the anti-correlation condition
$\sum_j\overline{h_ih_j} =0$. If this condition were not satisfied,
the properties would be the same as for uncorrelated fields. $c(\mu )$ 
vanishes as $\mu $ approaches zero. Therefore, the limit $\mu =0$
corresponds to uncorrelated fields.

Let us start with the case $T=0$. In order to obtain $d_l$ we have again to calculate

\begin{equation}
E^2_{\rm RF}\approx \sum\limits_{|{\bf r}_i-{\bf r}_j|<R}
\overline{h_ih_j}\sim h^2\left( R^{d-2\mu }+R^{d-1}\right)
\label{eq:E2RF}
\end{equation}
where the second term on the rhs is a surface contribution.
Comparing $E_{\rm RF}$ with the surface energy
$JR^{d-1}$ we obtain $d_l=2(1-\mu )$ if $\mu <1/2$.
For $\mu >1/2$ the random field does not affect the lower critical dimension
and hence $d_l=1$.

However, as we will show now, these are not the lower critical dimensions at
{\em finite temperatures}. To illustrate this point we consider random field
dimers, which consist of anti-parallel random fields $\pm h$
on two neighboring lattice sites and which are a particular realization
of anti-correlated random fields corresponding to large values of $\mu $.
Clearly, at $T=0$,  dimers
(compare Fig. 2) do not change the energy $E(M=\pm1)$ of the two ferromagnetic 
ground states with magnetization $M=\pm 1$
and hence $E(M=\pm1)=E_0$. 
\begin{figure}[h]
  \centerline{\psfig{figure=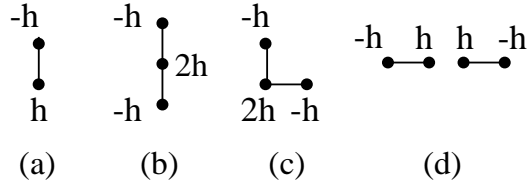,width=7cm}}
  \caption{Different configurations of random field dimers}.
\end{figure}
Random field
dimers act here similar to (oriented) random bonds, which
can lower the surface energy of a domain, as follows from the second term in
Eq. \ref{eq:E2RF}.

The situation becomes however different if we consider the free 
energy difference of the  states $M\approx\pm 1$ at non--zero temperatures
$0<T\ll h\ll J$. 
For a
single isolated dimer (Fig 2.a) we find from the low temperature expansion of
the free energy again no symmetry breaking effect
\begin{equation}
F(M\approx \pm 1)=E_0-T e^{-2zJ/T}\left( N-2+e^{2h/T}+e^{-2h/T}\right)
+O\left( e^{-4zJ/T}\right) . 
\label{eq:F(1)1}
\end{equation}
Here $z$ denotes the number of nearest neighbors of a spin.
The second term on the rhs of Eq. \ref{eq:F(1)1} arises from flipping a single spin only. On the other hand, for two dimers forming a pair as in
Fig. \,2b,c we get instead
\begin{equation}
F(M\approx \pm 1)=E_0-T e^{-2zJ/T}\left( N-3+e^{\mp 4h/T}+2e^{\pm 2h/T}
\right) +O\left( e^{-4zJ/T}\right) . 
\label{eq:F(1)2}
\end{equation}
Thus
\begin{equation}
F(1)-F(-1)\approx 2Te^{-2zJ/T}\left( \sinh{(4h/T)}-2\sinh{(2h/T)}\right)\equiv
2h_{\rm eff} 
\label{eq:heff}
\end{equation}
where $h_{\rm eff}$ can be considered as the effective random field strength
of the dimer pair. Similar symmetry breaking contributions follow
in higher order in $\exp{(-2zJ/T)}$ also from spatially separated
dimers as in Fig.\,2d. and from other realizations of
anti-correlated fields~\cite{SchVShN}. Thus anti-correlated random fields 
generate uncorrelated random fields, as can alternatively be 
seen from a perturbative 
treatment of the Ginzburg-Landau-Hamiltonian~\cite{SchVShN}.

In $d=2$ dimensions we come therefore to the surprising 
conclusion, that
although a system with anti-correlated fields 
has a ferromagnetically ordered ground state, ferromagnetic
order is destroyed at all non--zero temperatures. The
correlation length diverges for $T\rightarrow 0$ as 
$\exp{(J/h_{\rm eff})^2}$ where $h_{\rm eff}$ is given in Eq. \ref{eq:heff}.
We note finally, that anti-correlated random fields may play a role
in the description of binary fluids in Vycor~\cite{SchVShN}.

\subsection{Interface properties}\label{subsec:interfprop}

In the argument of Imry and Ma~\cite{IM} one assumes, that domain walls 
are smooth such that their surface area is proportional to $R^{d-1}$
if $R$ denotes linear extension 
of the domain. A closer inspection shows, that
this is indeed true in $d>2$ dimensions.
For simplicity we consider here a domain wall with a vanishing
mean curvature, which can be considered to be a small part of the surface
of a large domain. If we denote the interface profile  by $z(\underline{x})$ 
where $\underline{x}$ denotes a $(d-1)$-dimensional vector parallel 
to the mean interface plane and ${\bf r}=(\underline{x},z)$,
the interface Hamiltonian is 

\begin{equation}
{\cal H}_{\rm I}=\int{\rm d}^{d-1}x\,\left\{ \sigma_0+\frac{1}{2}\Gamma
(\underline{\nabla }\,z)^2-2M\int_0^{z(\underline{x})} {\rm d}z'\,
h(\underline{x},z')\right\}.
\label{eq:HI}
\end{equation}
Here $\sigma_0$ and $\Gamma$ denote the bare surface tension and 
the wall stiffness, respectively. 

Eq. \ref{eq:HI} can be derived in principle from the 
lattice Hamiltonian of the random field Ising model.
$\sigma_0$ and $\Gamma$ depend then in general 
on $J, h$ and $T$ in a complicated manner. 
At $T=0$, $\sigma_0\sim J$ and $\Gamma \sim J(J/h)^{2/(3-d)}$ for $d<3$ and 
$\Gamma \sim J e^{c(J/h)^2}$ for $d=3$ ~\cite{natter1983}.  
For $3<d<5$ dimensions and $h<h_c$  domain walls are flat because of 
the influence of the periodic 
potential of the underlying lattice, and hence cannot be 
described by Eq. \ref{eq:HI}. At $h_c$ the wall
undergoes a roughening transition such that for $h>h_c$  Eq. \ref{eq:HI} 
again applies~\cite{natter1983,DSF86a,bouchaud}. 
In the following we will assume, that we are always at $h>h_c$ .

The properties of the random field $h(\bf r)$ are still given by 
Eq. \ref{eq:barhr}.
The smoothness of the domain wall can be checked by considering
the interface roughness
\begin{equation}
w(L)=\overline{<(z(\underline{x}_1)-z(\underline{x}_2))^2>}^{1/2}\,
\Big|_{|\underline{x}_1-\underline{x}_2|=L}\sim L^{\zeta}.
\label{eq:wL}
\end{equation}
If the exponent $\zeta$ is nonzero, but less than one, the interface 
is termed {\it self-affine}.

A simple Imry--Ma argument for the energy of a hump of a linear extension
$L$ and height $w$ gives~\cite{grinma}
\begin{equation}
w(L)\approx \left(\frac{Mh}{\Gamma}\right)^{2/3}L^{(5-d)/3}\approx \left(\frac{L}{L_c}\right)^\frac{5-d}{3}.
\label{eq:wL1}
\end{equation}
The expression on the rhs of Eq. \ref{eq:wL1} defines the Larkin length $L_c$, to which we come back in Section \ref{subsec:0tempdep}.
This result Eq. \ref{eq:wL1} has been confirmed 
in $d=5-\tilde{\epsilon}$ dimensions by a functional renormalization group
calculation~\cite{DSF86a} and in $d=2$ by a mapping on the Burger's 
equation~\cite{Natter87}. The roughness exponent $\zeta =(5-d)/3$
is smaller than one in all dimensions $2<d<5$. For  $d>5$ 
dimensions domain walls are always flat (one could also say, 
that $h_c\rightarrow \infty$). Thus in all dimensions $d>2$, the 
typical gradients of the walls are small $\sim L^{\zeta -1}\sim L^{(2-d)/3}$
and hence walls are smooth on large scales. On the other hand,
in $d=2$ dimensions domain walls are rough on all scale, which 
leads to a vanishing total surface tension
on length scales 
$L \ge R_c\sim \exp(c \sigma_0\Gamma^{1/3}h^{-4/3})\sim \exp{[c'( J/h)^2]}$
 ~\cite{Bi}. $c$ and $c'$ are constants of order one.

\section{Phase Transition}\label{sec:phasetrans}
\subsection{Order of the transition}\label{subsec:orderoftrans}

In the last section we convinced ourselves, that the Ising model in a short 
range correlated random field has an ordered phase in more than two space 
dimensions. A natural question is then about the order of the
transition.

The mean field approximation, which is exact for the infinite--range model,
predicts a second order transition for a Gaussian distribution
of the random field strength. For a bimodal distribution the transition
however becomes first order if $h$ is larger than a critical value~\cite{Aha}.
For a model with short range interaction earlier numerical 
simulations~\cite{YoungN} and high temperature expansions~\cite{Khurana}
seem to confirm a first order transition, but now independent
of the field distribution.

More recent Monte Carlo simulations for bimodal and Gaussian random
field distributions, however, do not show a latent heat~\cite{RY} or 
phase coexistence~\cite{R}, 
but a jump in the magnetization $M$. The latter may be related
to a logarithmic dependence of $M$ on the reduced temperature.
Also recent high temperature series expansion up to 15 terms shows a 
continuous transition for both
distributions~\cite{Gof}. On the other hand, Swift et al.~\cite{Swift} find 
in a recent study from an exact determination of the ground states 
in higher dimensions ($d=4$) 
a discontinuous transition for a bimodal field distribution whereas for
a Gaussian distribution and in $d=3$ dimensions the transition is continuous.

Although there is so far no proof for a continuous transition in neither 
case, we will
assume in the following that the transition is indeed continuous, which seems 
to be at least true for a Gaussian random field distribution.

\subsection{Scaling laws}\label{subsec:scalinglaws}

Under which conditions are random fields relevant at a critical point?
To answer this question we consider the system at the true transition 
temperature $T_c(h)$ 
in a finite 
{\em homogeneous} field $H$. We divide the system into blocks of linear size
$\xi$. Here $\xi$ denotes the correlation length which is related
to $H$ by $H\approx J\xi^{-\beta\delta /\nu}$. Here we follow the standard 
notation for the critical exponents $\alpha, \beta, \gamma, \delta, \nu$ etc.~\cite{Stanley}. The random field produces
an additional excess field $\delta H\approx ch\xi^{-d/2}$
where $c$ is a constant of order one and arbitrary sign.
Approaching the critical point $H\rightarrow 0$ we expect a sharp 
transition,
 if for $\xi\rightarrow\infty$, 
$\delta H(\xi )/H(\xi )\rightarrow 0$,
i.e. for
\begin{equation}
\frac{d}{2}\ge\frac{\beta\delta}{\nu}=\frac{2+\gamma-\alpha}{2\nu}.
\label{eq:d/2}
\end{equation}
On the rhs of Eq. \ref{eq:d/2} we used a $d$--independent scaling law
which is assumed to be valid also for random field models. 
With the mean--field
exponents $\beta =\nu =1/2,\,\,\delta =3$ we find, that random
field effects are negligible only for $d\ge d_c=6$. Thus
random fields are relevant for $d\le 6$ ~\cite{grinstein}.
For $d<4$, where the pure model has already non--classical exponents
$\alpha^{(0)}$, $\nu^{(0)}$ etc. we get from eq. Eq. \ref{eq:d/2}, 
and using 
the {\em conventional} hyper-scaling law 
$\nu^{(0)}d=2-\alpha^{(0)}$ the relation $\gamma^{(0)}\le 0$. Since the 
susceptibility exponent $\gamma^{(0)}$ has to be positive, we have to 
conclude, that the random field is also a relevant
perturbation if we use the critical exponents of the pure model and moreover, 
that the conventional hyperscaling breaks down in random field systems.

The condition for the applicability of the linear response theory, 
$\delta H(\xi )\le H(\xi )$, is violated if

\begin{equation}
\xi\ge\xi_{\rm LG}\approx \left(\frac{J}{h}\right)^{2\nu /(2\beta\delta
-\nu d)}.
\label{eq:xi0}
\end{equation}
Eq. \ref{eq:xi0} represents the Levanyuk--Ginzburg critical 
region, where random field
effects cannot be ignored, since different volumes perceive different
effective fields $H+\delta H$. The resulting transition could be 

\noindent (i) {\em smeared} (as one could naively expect) or 

\noindent (ii) {\em 1st order}
(as ruled out in the previous paragraph), or 

\noindent (iii) {\em 2nd order} with
{\em modified exponents} and {\em modified scaling relations} (because
we would get the unphysical result $\gamma\le 0$ with conventional hyper-scaling relation). 

Let us assume that case (iii)  (as simulations suggest) and the  
modified hyper-scaling relation~\cite{VII,DSF86b}
\begin{equation}
\nu (d-\theta )=2-\alpha
\label{eq:nu}
\end{equation}
with some new exponent $\theta$ apply. Then relation Eq. \ref{eq:d/2} 
leads to the condition
\begin{equation}
\theta\ge\gamma /\nu =2-\eta
\label{eq:theta0}
\end{equation}
an inequality which has been proven exactly by Schwartz and Soffer for
a number of random field systems~\cite{SchwaSoff1}.
 
A heuristic argument to make Eq. \ref{eq:nu}
obvious follows the original idea of Pippard: the free energy of a correlated 
region $F_{\xi}$ scales as the energy, which is necessary to flip the region.
In pure systems this energy is $k_BT$ whereas in random field
systems it can be estimated as $h\xi^{\theta}$. 
Our previous Imry--Ma argument would give $\theta\approx d/2$, but we have to 
take into account, that the random field gets renormalized
close to criticality. Thus
\begin{equation}
F_{\xi}\approx J\xi^{d-(2-\alpha )/\nu}\sim k_BT+h\xi^{\theta}
\label{eq:xi}
\end{equation}
from which we can immediately read off Eq. \ref{eq:nu}. 
$\theta$ is a third independent 
critical exponent that describes the irrelevance of temperature
as is plausible already from Eq. \ref{eq:xi}.

The appearance of a third exponent is related to the existence of
two different correlation functions which scale independently. In particular, at $T=T_c(h)$
\begin{equation}
G({\bf k})=\overline{\left<\tilde{S}_{\bf k}\tilde{S}_{-{\bf k}}\right> }
-\overline{\left< \tilde{S}_{\bf k}\right> \left< \tilde{S}_{-{\bf k}}\right> }
\sim k^{-2+\eta}
\label{eq:Gk}
\end{equation}
and
\begin{equation}
C_{dis}({\bf k})=\overline{\left< \tilde{S}_{\bf k}\right> \left< 
\tilde{S}_{-{\bf k}}\right> }\sim k^{-4+\bar{\eta}}
\label{eq:Ck}
\end{equation}
where $\tilde{S}_{\bf k}=\frac{1}{\sqrt N}\sum\limits_i
\exp{(i{\bf k}{\bf r}_i)}S_i$ and~\cite{VII,DSF86b}
\begin{equation}\bar{\eta}=2+\eta -\theta \,.
\label{eq:bareta}
\end{equation}
Note, that $C_{dis}({\bf k})$ vanishes for pure systems.

Besides the Schwartz--Soffer inequality there is a number of further 
inequalities. From $\beta\ge 0$ follows with Eq. \ref{eq:nu} and
Eq. \ref{eq:bareta} $\bar{\eta}\ge 4-d$. Further inequalities are
$\gamma\ge 0$ ~\cite{VII} and $d-1\ge \theta $ ~\cite{DSF86b}.
The latter inequality follows from the fact, that the domain wall
energy scales as $\xi^{-\mu /\nu}$ with $\mu =\nu (d-1-\theta )\ge 0$.
Thus, the values of $\beta /\nu$ and $\theta $ are restricted to a pentagon
(see Fig. 3). Schwartz and Soffer have further claimed that 
Eq. \ref{eq:theta0} is fulfilled as an equality~\cite{SchwaSoff2}. 
Finally, there is 
the Harris-inequality~\cite{Chayes} $\nu \ge 2/d$.

\begin{figure}[h]
 \centerline{\psfig{figure=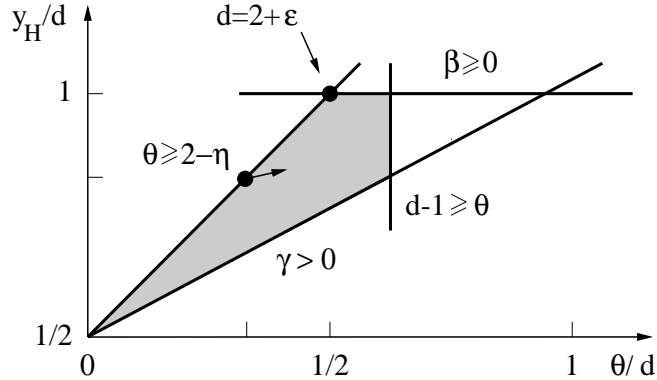,height=2.0in}}
 \caption{The domain of allowed values of $y_H=d- \beta /\nu $ and $y_J=\theta $. The arrow denotes the result of the perturbative expansion in $d=6-\epsilon $ dimensions. Most of the results for the exponents found numerically are located in the upper left corner }
\end{figure}

\subsection{Renormalization group}\label{subsec:renormgr}

Since $\theta >0$, it is plausible from Eq. \ref{eq:xi}, that in the 
critical region thermal fluctuations are less relevant than those of the 
frozen--in disorder. In a renormalization--group (RNG)
treatment this feature is reflected in the existence of a $T=0$
fixed point, which is believed to describe the critical behavior
of the random field Ising model.

Although there is so far no satisfying RNG analysis, we present
here briefly a rough sketch of it, assuming a continuous
transition up to $T=0$. We follow thereby  closely the earlier work 
of Bray and Moore~\cite{Bray}.
We start with the observation, that the free energy density $f$
can be written in the form $f = J {\tilde f}(T/J, h/J, H/J)$. Let us 
imagine to 
carry out the RNG coarse--graining transformation, with length scale
factor $b$, corresponding to a reduction in the number of degrees of freedom
by a factor $b^d$. The transformation generates a flow in the space of the 
naive scaling fields $T/J,\,\,h/J$, and $H/J$, which eventually terminates 
in one of the fixed points of the system. The existence of three fixed point 
will be assumed (Fig. 4), in addition to the trivial, high temperature
fixed point:

\noindent (i) A totally unstable ``thermal'' fixed point $C$ at $T=T_c$,
$H=h=0$ (the random field is a relevant perturbation, see our discussion
in \ref{subsec:scalinglaws}) .

\noindent (ii) A fixed point $R$ at $T=H=0$ and $h=h_R$ which is unstable
in two, but stable in one directions and is therefore a critical point.

\noindent (iii) A totally stable fixed point $F$ at $T=h=H=0$, which corresponds
to the low temperature phase for $d>d_l$.

\begin{figure}[h]
  \centerline{\psfig{figure=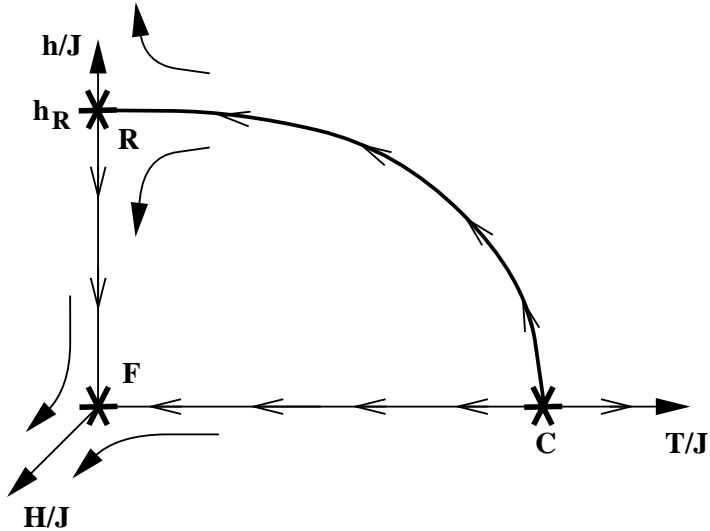,width=10cm}}
  \caption{Schematic renormalization group flow of the random field Ising model }
\end{figure}

In general the RNG procedure generates also new terms in the 
Hamiltonian. We will assume, that these terms are irrelevant in the 
RNG--sense and can therefore be neglected.

In order to calculate the critical behavior  
we have to linearize the 
RNG flow close to the fixed point $R$. The eigenvalues and eigenvectors of 
the linearized RNG--transformation deliver  the critical
exponents and scaling fields. Phenomenological arguments 
concerning the RNG flow suggest
\begin{equation}
\frac{T}{J},\quad \tau=\frac{1}{J}(h-h_R)+c\frac{T}{J}\quad {\rm and}\quad
\frac{H}{J}
\label{eq:T/J,t,H/J}
\end{equation}
as the scaling fields. Close to the fixed point $R$ \quad $J$, $\tau$ and $H$
transform under the RNG coarse graining as
\begin{equation}
\begin{array}{ccc}
J\rightarrow J' & = & J\,b\,^{y_J}\\
\tau \rightarrow \tau' & = & \tau\,b\,^{y_{\tau}}\\
H\rightarrow H' & = & H\,b\,^{y_H}.
\label{eq:JtH}
\end{array}
\end{equation}

From Eq. \ref{eq:T/J,t,H/J}, Eq. \ref{eq:JtH} and the invariance of the 
partition function under the RNG transformation we get for the 
singular part of the free energy density $f=J\cdot {\tilde f}$
\begin{equation}
{\tilde f}\left( \frac{T}{J},\tau ,\frac{H}{J}\right) =b^{y_J-d}{\tilde f}\left( \frac{T}{J}
b^{-y_J},\tau b^{y_{\tau}},\frac{H}{J} b^{y_H-y_J}\right)
\label{eq:Jf}
\end{equation}
and similar for the correlation length $\xi$
\begin{equation}
\xi \left(\frac{T}{J},\tau ,\frac{H}{J}\right) =b\xi \left(\frac{T}{J} b^{-y_J},
\tau  b^{y_{\tau}},\frac{H}{J} b^{y_H-y_J}\right).
\label{eq:xi1}
\end{equation}
The critical exponents follow from Eq. \ref{eq:Jf} and Eq. \ref{eq:xi1} in
the usual way, their relations to eigenvalue exponents $y_i$ are
summarized in the following list:
\begin{equation}
\begin{array}{l@{\qquad}l@{\qquad}l}
\beta =\nu (d-y_H) & \nu =1/y_{\tau} & \theta =y_J\\
\gamma =\nu (2y_H-y_J-d) & \eta =2+d+y_J-2y_H &\\
\delta =(y_H-y_J)/(d-y_H) & \bar{\eta}=4+d-2y_H  &.
\label{eq:list}
\end{array}
\end{equation}
These scaling relations are in 
agreement with Eq. \ref{eq:nu} and Eq. \ref{eq:bareta}. 
The existence of a zero--temperature
fixed point corresponds to a positive value of $y_J\equiv \theta$.
Alternatively, $-y_J$ can be interpreted as the eigenvalue exponent
of the temperature or, by simple scale transformation $\phi =\sqrt{T\phi '}$
in Eq. \ref{eq:HGL}, of the coupling constant $u$. With this interpretation of 
$-y_J$, the modified hyper-scaling law Eq. \ref{eq:nu} has been first derived by 
Grinstein~\cite{grinstein}. The RNG program scetched above 
has been performed in real space using Migdal-Kadanoff or related 
approximations 
by a number of 
groups~\cite{DaySchY,New,CaoMach,FalBMcK} (see Table \ref{tab:critexp}).

\subsection{Dimensional reduction and weighted mean field
approximation}\label{subsec:dimred}

A convenient starting point to study the critical behavior is the 
Ginzburg--Landau Hamiltonian ${\cal H}_{\rm GL}\{\phi\}$. 
Including a random field term, Young~\cite{Y} was
able to show that the most singular terms in the 
perturbation theory for 
this model follow then from  tree diagrams, which can alternatively be 
obtained from an iterative solution of the saddle
point equation
\begin{equation}
\delta {\cal H}_{\rm GL}/\delta \phi =
-{\bf \nabla}\phi^2+V'(\phi )-h({\bf r})=0.
\label{eq:nablaphi}
\end{equation}
Here $V(\phi )=\frac{1}{2}{\tilde a}\phi^2+\frac{1}{4}u\phi^4$.
The contributions from tree diagrams lead to an exact relation
between the critical exponents $\lambda_i^{({\rm RF})}(d)$ of
the random field system in $d$--dimension and those of the
pure model in $(d-2)$--dimensions
\begin{equation}
\lambda_i^{({\rm RF})}(d)=\lambda_i^{({0})}(d-2)
\label{eq:lambdaRF}
\end{equation}
the so called {\em dimensional reduction}.
This equality has been proven, starting from Eq. \ref{eq:nablaphi}, also in
a non--perturbative way~\cite{ParS}. Comparing Eq. \ref{eq:lambdaRF}
with eq. Eq. {\ref{eq:nu} yields $\theta =2$ in all dimensions. Since the lower
critical dimension of the pure Ising model is $d_l^{({0})}=1$
one would conclude $d_l^{({\rm RF})}=3$, in disagreement with our findings
from the previous Section.

Apparently, perturbation theory is inappropriate to deal with this type
of disorder which is characterized by a large number of local
minima in the energy landscape. Indeed, since the perturbation
theory can be generated from Eq. \ref{eq:nablaphi}, all saddle point
solutions enter expectation values of physical quantities with
the weight $\pm 1$, which is clearly the wrong way.

More recently Lancaster et al.~\cite{LMP} proposed a weighted mean 
field theory
which takes into account {\em all} solutions $m_i^{\alpha}$ of the mean field 
equations
\begin{equation}
\partial F/\partial m_i=-J{\left.\sum\right._j}^{(i)}m_j+T\,{\rm arctanh}\, m_i-h_i=0, \qquad i=1,...,N .
\label{eq:meanfieldeq}
\end{equation}
Here ${\left.\sum\right._j}^{(i)}$ denotes the sum over the nearest neighbors
to the site $i$. $h_i$ is taken from  a bimodal distribution.
Thermodynamic quantities are then calculated as a sum over all mean
field solutions $\alpha$ with the Boltzmann weight
$\exp{(-F_{\alpha}/T)}/\left.\sum\right._{\alpha}\exp{(-F_{\alpha}/T)}$
where $F_{\alpha}=E_{\alpha}-TS_{\alpha}$ is the mean field free energy and 
\begin{equation}
\begin{array}{rcl}
& & E_{\alpha} = -\frac{1}{2}J\sum\limits_{<ij>}m_i^{\alpha}m_j^{\alpha}
-\sum\limits_ih_im_i^{\alpha}\\
& & \\
& & S_{\alpha} = -\sum\limits_i\frac{1}{2}\left\{ (1+m_i^{\alpha})
\ln{(1+m_i^{\alpha})}+(1-m_i^{\alpha})\ln{(1-m_i^{\alpha})}-2\ln 2\right\}
\end{array}\label{eq:freeenergy}.
\end{equation}
We note, that Eq. \ref{eq:meanfieldeq} is a lattice version of
Eq. \ref{eq:nablaphi} with $V(m_i)=-T_cm_i+T\,{\rm arctanh}\, m_i$
and $T_c=2{\rm d}J$.
Since Eq. \ref{eq:meanfieldeq} represents the saddle points of 
$F_{\alpha}$ and thermal fluctuations are expected to be irrelevant
for the critical behavior, in principle one should be able 
to calculate the true critical exponents from this approach (at least at 
sufficiently low temperatures).
Starting from high temperatures $T>T_c^{({\rm pure})}$ there is typically only
one solution to the mean field equation. Decreasing $T$
below a temperature $T^{\ast}>T_c$, the number of solutions starts to grow 
rapidly.
For a $32^3$ periodic lattice and an accuracy
$|m_i^{\alpha}-m_i^{\gamma}|\stackrel{>}{\sim} 10^{-3}$, this 
number becomes of the order 250, although the number of those with
a large weight increases more slowly. The critical exponents obtained in
this way are summarized in Table~{\ref{tab:critexp}. Lancaster et al.~\cite{LMP} identify $T^{\ast}$ with the temperature $T_{RSB}$, at which replica 
symmetry breaking occurs (see Section 3.6).

\subsection{Two or three independent exponents ?}{\label{subsec:2or3exp}

In contrast to conventional critical points, which are characterized
by two independent critical exponents, our schematic renormalization group
calculation suggests, that the random field Ising model is
characterized by three independent exponents.

In an early publication, Aharony, Imry and Ma~\cite{AhaIM} suggested
the existence of an exact exponent relation
\begin{equation}
\theta =2-\eta
\label{eq:theta}
\end{equation}
which implies $2\eta =\bar{\eta}$ and reduces the number of independent 
exponents again to two. Relation Eq. \ref{eq:theta} can indeed be made plausible by 
estimating the free energy of a correlated droplet $F_{\xi}$
close to the critical point. With 
$M({\bf r})\approx \chi \, h \xi^{-d/2}\sim h\xi^{2-\eta -d/2}$
for the local magnetization, where $\chi =G({\bf 0})/T $ denotes the susceptibility, and Eq. \ref{eq:xi} we get
\begin{equation}
F_{\xi }\sim\int
_{{\bf r}\in\xi^d}{\rm d}^dr\,h({\bf r})M({\bf r})\sim
h^2\xi^{2-\eta}\sim\xi^{\theta}
\label{eq:Fxi}
\end{equation}
which gives Eq. \ref{eq:theta}. 
Later Schwartz et al.~\cite{SchwaSoff2} have claimed, that
there is an exact proof for this relation. However, all these approaches 
use in one or the other way linear response arguments and have  therefore 
to be considered with caution. Numerical calculations
show, that Eq. \ref{eq:theta}  is indeed fulfilled within 
the accuracy of the calculation.

Since the numerical determination of exponents in random systems
is typically hampered by the existence of considerable error bars,
which makes a confirmation of Eq. \ref{eq:theta} difficult, Gofman et 
al.~\cite{Gof} 
considered the even stronger relation
\begin{equation}
A=\lim\limits_{T\rightarrow T_c}\frac{T^2}{h^2}\frac{C_{dis}({\bf 0})}{G^2({\bf 0})}=1
\label{eq:A}
\end{equation}
which should be fulfilled according to~\cite{SchwaSoff2}.

The exponent scaling gives for $C_{dis}(0)/G^2(0)\sim\xi^{2\eta -\bar{\eta}}$,
which would diverge (compare Eq. \ref{eq:theta0} and 
Eq. \ref{eq:list}) unless
Eq. \ref{eq:theta} is valid. Gofman et al. determine $A$ from a 15 terms
high temperature series expansion for $G(0)+C_{dis}(0)$ and
$G(0)$ in $d=3,\,4,\,5$ dimensions for different values of the
random field strength $h$. In particular, for $d=3$, $A=1\pm 0.003$
which is an impressive confirmation of Eq. \ref{eq:A} and Eq. \ref{eq:theta}
(but not a proof!). Thus, despite of all efforts to prove Eq. \ref{eq:theta}
this problem has still to be considered as unsolved.

\begin{table}[t]
\caption{Critical exponents.\label{tab:critexp}}
\vspace{0.4cm}
\begin{center}
\small
\begin{tabular}{|l|c|c|c|c|}
\hline
\rule[-2mm]{0mm}{0.7cm}
Reference & $\theta =y_J$ & $\nu =1/y_{\tau}$ & $\beta =\nu (d-y_H)$ &
$2\eta -\bar{\eta}=\theta-2+\eta $\\\hline\hline
\rule[0mm]{0mm}{0.5cm}
${\rm MFA}^{\,\,}$\footnotemark[1]
& $d-4$ & 1/2 & $1/2$ & $d-6$\\\hline
\rule[0mm]{0mm}{0.5cm}
${\rm PT}^{\,\,}$\footnotemark[2] $d=6-\epsilon $ 
& 2 & $\nu^{({0})}(d-2)$ & $\beta^{({0})}(d-2)$ 
& $\eta^{({0})}(d-2) $
\\\hline
\rule[0mm]{0mm}{0.5cm}
$d\rightarrow 2$ & 1 & 0 & 2 & 0\\\hline
\rule[0mm]{0mm}{0.5cm}
weighted MFA~\cite{LMP} & 1.51 & $1.25\pm 0.11$ & &\\\hline
\mco{1}{|l|}{\rule[0mm]{0mm}{0.5cm}
${\rm {\bf MC}}^{\,\,}$\footnotemark[3] (d=3)} & & & &\\
{\scriptsize Rieger and Young~\cite{RY}} & $1.56\pm 0.1$ & $1.6\pm
0.3$ &$0.003\pm 0.05$ & $0.12\pm 0.12$\\
{\scriptsize Rieger~\cite{R}} & $1.53\pm
0.1$ & $1.1\pm 0.2$ & $0.00\pm 0.05$ & $-0.03\pm 0.15$\\
\hline

\begin{minipage}{1in}
\rule[0mm]{0mm}{0.5cm}
${\rm HTSE}^{\,\,}$\footnotemark[4] \\ \rule[-2mm]{0mm}{0.5cm}
{\scriptsize Gofman et al.}~\cite{Gof}\end{minipage} & 
$(\gamma =2.1\pm 0.2)$& & & 0\\\hline

\mco{1}{|l|}{\rule[0mm]{0mm}{0.5cm}
Realspace RNG} & & & &\\ 
\begin{minipage}{1in}
{\scriptsize Dayan et al.}~\cite{DaySchY}\\ 
{\scriptsize Newmann et al.}~\cite{New}\\
{\scriptsize Cao and Machta}~\cite{CaoMach}\\ 
{\scriptsize Falicov et al.}~\cite{FalBMcK}\rule[-2mm]{0mm}{0cm}
\end{minipage}& 
\begin{minipage}{1in}
1.56\\ $1.00\pm 0.05$\\
1.5\\ $1.4916\pm 0.0003$\end{minipage}
& \begin{minipage}{1in} 
$1.39$\\ $1.49\pm 0.008$\\ $\nu =2.25$\\ $\nu =2.25\pm 0.01$
\end{minipage}& 
\begin{minipage}{1in} $\approx 0$\\ $1.66\pm 0.01$\\ 0.02\\ $0.0200\pm
0.0005$\end{minipage}& 
\begin{minipage}{1in} 0.12\\ 1.24\\
0.02\\ $0.001\pm 0.001$ \end{minipage}\\\hline
\end{tabular}

\end{center}
\end{table}
\footnotetext[1]{MFA: mean field approximation} 
\footnotetext[2]{PT: perturbation theory}
\footnotetext[3]{MC: Monte Carlo simulation}
\footnotetext[4]{HTSE: high temperature series expansion}

\subsection{Replica symmetry breaking (RSB)}\label{subsec:RSB}

The disorder average can conveniently be performed using the replica trick. 
E.g. the average free energy can be written as
\begin{equation}
\bar{F} =-T\overline{\ln Z}=-T\lim\limits_{n\rightarrow 0}
\frac{1}{n}(\overline{Z^n}-1)
\label{eq:avfreeen}
\end{equation}
where $\overline{Z^n}=\overline{({\rm Tr}\,\exp{(-{\cal H}/T)})^n}={\rm Tr}\,
\exp{(-(1/T)
{\cal H}_n\{ S_i^{\alpha}\})}$ is the conventional partition function
of the translationally invariant replica Hamiltonian
\begin{equation}
{\cal H}_n\{ S_i^{\alpha}\} =-\sum\limits_{\alpha ,\beta}\sum\limits_{i,j}
\left( J_{ij}\delta_{\alpha\beta}+\frac{h^2}{2T}\delta_{ij}\right) S_i^{\alpha}S_j^{\beta}.
\label{eq:HnS}
\end{equation}
An alternative formulation uses the Ginzburg--Landau model as
the starting point with the corresponding replica Hamiltonian
\begin{equation}
{\cal H}_{GL,n}\{ \phi^{\alpha}({\bf r})\} =\sum\limits_{\alpha}{\cal H}_{\rm GL}
\{ \phi^{\alpha}\} -\sum_{\alpha,\beta}\int {\rm d}^dr\frac{h^2}{2T}\phi^{\alpha}\phi^{\beta} .
\label{eq:Hnphi}
\end{equation}
Mezard and Young~\cite{MY} use an $m$--component generalization 
of Eq. \ref{eq:Hnphi}, $\phi^{\alpha}({\bf r})\rightarrow 
 \vec {\phi }^{\alpha}({\bf r})= (\phi_i^{\alpha}(r), i=1\ldots m)$, 
to determine the correlation 
functions $G^{\alpha\beta}({\bf k})$ and $C_{dis}^{\alpha\beta}({\bf k})$
from Bray's self--consistent screening approximation (SCSA). Note, that 
this generalization changes the lower critical dimension from $d_l=2$ to 
$d_l=4$, which follows both from perturbation theory and, because of the 
continuous symmetry of the order parameter, 
also from the Imry-Ma argument~\cite{IM}. However, 
dimensional reduction is expected to break down also in 
this case~\cite{DSF85}.

The SCSA is a truncation of Dyson's equation which is exact to
order $1/m$. Assuming a replica--symmetric solution, Mezard and Young
find the dimensional reduction result 
$\eta (d)=\bar{\eta}(d)=\eta ^{(0)}(d-2)=O(1/m)$. 
This replica symmetric solution 
is however unstable with respect to RSB. Using the hierarchical RSB scheme 
of Parisi, they find 
$\eta =\eta^{(0)}(d-2)$ to order $1/m$, but the value $\bar{\eta}$
is now altered. The difference $2\eta -\bar{\eta}=c/m$
is small for any $m$.
The constant $c$ can in principle be determined from the set of 
self--consistent equations. This calculation has been extended by 
Mezard and Monasson~\cite{MM} and de Dominicis et al.~\cite{DOT},
who determined the temperature $T_{\rm RSB}>T_c$, where the replica 
symmetric solution becomes unstable. In particular, de Dominicis 
et al.~\cite{DOT} consider a more general coupling term
$\left.\sum\right._{\bf k} \Delta_{\alpha\beta}({\bf k})\phi_{\bf k}^{\alpha}
\phi_{-{\bf k}}^{\beta}$ between the replicas in Eq. \ref{eq:Hnphi}
and determine $T_{\rm RSB}$ from the divergence of the $\Delta$--susceptibility
$\partial^2\overline{Z^n}/\partial\Delta_{\alpha\beta}({\bf p})
\partial\Delta_{\gamma\delta}({\bf p}')$
(to reach this goal, in practice they use the 
Legendre--transform of $\overline{Z^n}$), which is related to 
the standard spin--glass susceptibility. It turns out, that 
the spin--glass transition (in the sense of an Almeida-Thouless line),
which is believed to take place at $T_{\rm RSB}$, always precedes the 
ferromagnetic transition (see Fig. 5). For $d$  close to six dimensions
they get in particular
\begin{equation}
|T_{\rm RSB}-T_c(h)|\approx J (h/J)^{8/(6-d)}.
\label{eq:Trsb-Tc}
\end{equation}
This has to be compared with the Levanyuk--Ginzburg criterion, Eq.\ref{eq:xi0},
which gives a random field controlled critical region of size
\begin{equation}
|T_{\rm LG}-T_c(h)|\approx J (h/J)^{4/(6-d)}\gg |T_{\rm RSB}-T_c(h)|.
\label{eq:Tlg-Tc}
\end{equation}
For $T_c<T<T_{\rm LG}$ we expect, that the saddle point equations
Eq. \ref{eq:nablaphi}, Eq. \ref{eq:meanfieldeq}, will have several solutions,
which signals the failure of perturbation  and linear
response theory. It is at present unclear, whether RSB (and hence the
breakdown of dimensional reduction) occurs in  the whole random field 
critical region, as one would naively expect and as was found in a more 
recent study by Dotsenko and 
Mezard~\cite{dotsenko}, or, as Eq. \ref{eq:Trsb-Tc} suggests, 
is restricted to 
the much smaller temperature interval around $T_c(h)$ ($h\ll J$). 
We note finally, that the spin-glass order parameter $q_{EA}$ is non-zero 
at all temperatures. In particular, outside the critical region where 
linear response theory applies 
$\overline{<S_l>^2}=q_{EA}\approx  \left.\sum\right._{ij} \chi_{li}\chi_{lj}
\overline{h_i h_j} = h^2 \left.\sum\right._{i}  \chi_{li}^2 >0$. 
Here $T\chi_{ij}=<S_i S_j>-<S_i><S_j> $ denotes the Fourier transform of $G({\bf k})$.

Although we have here 
only discussed the paramagnetic phase, one expects a similar behavior if 
one reaches the ferromagnetic transition line from below.

\section{Dynamical Properties}

\subsection{Zero temperature interface depinning}\label{subsec:0tempdep}
 
At finite temperatures the two phases with up and down
magnetization can coexist only for vanishing strength of the 
uniform external field $H=0$. This is not the case at
zero temperature, where the disorder (and possibly also lattice effects) 
lead to a pinning of the
wall separating the two domains. In order to get the wall depinned, the
external field has to overcome a threshold $H_c$. Thus, the coexistence 
surface consists of two parts, for $0<T<T_c(h)$ it is
restricted to $H=0$, whereas for $T=0$ it is given by $-H_c(h)\le H\le H_c(h)$ (see Fig. 5).

\begin{figure}[htb]
  \centerline{\psfig{figure=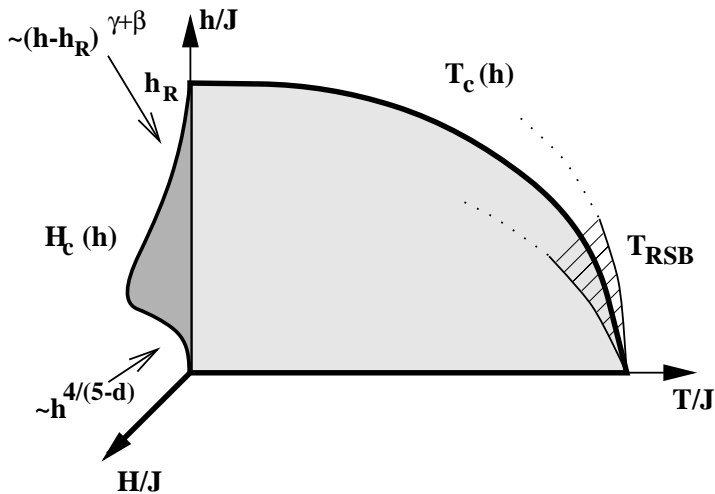,width=10cm}}
  \caption{Coexistence surfaces of the random field Ising model }
\end{figure}

In this section we consider the behavior of the interface in the
vicinity of the depinning threshold and determine $H_c$.
The equation of motion of an over-damped wall is~\cite{feigel} 
\begin{equation}
\lambda \frac{\partial z}{\partial t}=-\delta H_I/\delta z=
\Gamma\underline{\nabla}^2\,z+2M(H+h(\underline{x},z)).
\label{eq:lambdazt}
\end{equation}
This equation is highly non-linear because of the last term on the rhs.
$\lambda $ denotes the inverse mobility of the interface. We assume here, that the 
interface motion is over-damped.
We will show later, that an inertial term is indeed irrelevant close to 
the depinning threshold. 
It turns however out, that it is important to assume
a finite correlation length $a$ for the random field in the $z$--direction
(which is the direction of motion of the interface), i.e. we replace Eq. \ref{eq:barhr} by 
\begin{equation}
\overline{h(\underline{x},z)h(\underline{x}',z')}=h^2 \delta^{\rm d-1}
(\underline{x}-\underline{x}')\Delta (z-z')
\label{eq:hxz}
\end{equation}
with $\int_{-\infty}^{\infty}\Delta (z)=1$.
$a$ is of the order of the lattice spacing, i.e. of order one in our units.
$\Delta (z)=-\Delta(-z)$ is a monotonically decreasing function of 
$z$ for $z>0$
and decays rapidly to zero over a finite distance $a$. The
width of the random field correlator perpendicular to $z$ turns out to be 
irrelevant, we assume therefore, that it is smaller than any other
length of the problem.

The physics of the interface close to the depinning 
transition is characterized by two emerging important length scales,
the Larkin length $L_c$ and the dynamical correlation length
$\xi_{v}$. For weak disorder, $J\gg h$, $L_c$ is the length scale on 
which the typical distortion of the interface is of the order $a$. 
We have  to compare here the curvature force
$\Gamma L^{{\rm d}-3}a $ with the random force 
$\sim M h\Delta^{1/2} (a)L^{{\rm (d-1)}/2}$ .
In the case of weak disorder,
which we consider here, $\Gamma a\gg M h\Delta^{1/2}$, and 
the random force wins over the curvature force only on length scales
\begin{equation}
L>L_c\approx [(\Gamma a/Mh)^2/\Delta (a)]^{1/(5-{\rm d})}.
\label{eq:L>Lc}
\end{equation}
and ${\rm d}<5$.
For $L<L_c$ (or ${\rm d}>5$), the interface is flat and hence cannot be
pinned, since the total pinning force $\sim L^{{\rm (d-1)}/2}$ is 
always smaller than the driving force $\sim L^{\rm (d-1)}$. On
the other hand, the interface is able to explore the inhomogeneous
force field on larger length scales $L>L_c$. It follows, that the
maximum pinning force on a piece of interface of linear dimension $L>L_c$
is of the order $(L/L_c)^{\rm (d-1)}Mh[\Delta (a)L_c^{\rm (d-1)}]^{1/2}$,
which leads to a threshold field of the order
\begin{equation}
H_c\approx\frac{1}{M}\Gamma L^{-2}_ca\approx\frac{\Gamma a}{M}
\left[ \left(\frac{hM}{\Gamma a}\right)^2\Delta (a)\right]^{2/(5-{\rm d})}.
\label{eq:Hc}
\end{equation}

For small $h$, $H_c\sim h^{4/(5-{\rm d})}$, i.e. $H_c\sim h^2$
in $d=3$ dimensions. Close to the critical point $h\stackrel{<}{\sim }h_R$,
$H_c\approx\Gamma\xi^{-1}/M\sim \tau^{\beta\delta}$. Note however, that 
$\Gamma$ carries its own $h$-dependence if Eq. \ref{eq:lambdazt} 
is derived from the bulk 
Hamiltonian (compare Section 2.3). Then for $h\rightarrow 0$, 
$H_c$ becomes exponentially small in $d=3$ dimensions.

To understand the nature of $\xi_{v}$, it is convenient to describe 
the interface in a co--moving frame
$z(\underline{x},t)=h(\underline{x},t)+vt$ where $v=<\dot{z}>$ denotes the mean interface velocity. Hence
\begin{equation}
\lambda\frac{\partial h}{\partial t}=\Gamma\underline{\nabla}^2\,h+
2MH-\lambda v+2Mh(\underline{x},vt+h).
\label{eq:mu-1}
\end{equation}
Let us consider the motion of a domain wall over some typical
obstacle formed by a random field cluster, which is assumed to be 
hit by the wall
at $t=0$. For small $t$, the wall will locally be stopped 
 whereas other parts will continue to move forward by a distance
$vt$. If the moving wall behaves self--affine under the action of the random field
(an assumption which we will prove later), it will form locally a bump of 
typical height $h\sim t^{\tilde{\zeta} /\tilde{z}}$, where $\tilde{\zeta}$
and $\tilde{z}$ are the non--equilibrium roughness and the
dynamical index of the interface, respectively.
Since $\tilde{\zeta}<\tilde{z} $, on time scales 
$t\gg t_{v}\sim v^{-\tilde{z}/(\tilde{z}-\tilde{\zeta})}$ 
we find $ vt\gg h$ and hence the
non--linearity in Eq. \ref{eq:mu-1} can be neglected. 
On the time scale 
$t_{v}$ the local perturbation by the obstacle spreads out on the interface 
over a length scale

\begin{equation}
\xi_{v}\sim t_{v}^{1/\tilde{z}}\sim v^{-1/(\tilde{z}-\tilde{\zeta})}
\label{eq:locper}
\end{equation}
which has apparently the meaning of a dynamical correlation length. Thus,
on time and length scales $t\gg t_v$ and $L\gg \xi_v$,
 the effective interface equation is the linear 
Edwards-Wilkinson-equation~\cite{edwards}
\begin{equation}
\lambda_{\rm eff}\frac{\partial z}{\partial t}=\Gamma_{\rm eff}\underline{\nabla}^2z+
2M(H_{\rm eff}+h_{\rm eff}(\underline{x},vt))\label{eq:edwileq}
\end{equation}
but with renormalized parameters.
In particular, we will find $H_{\rm eff}=H-H_c$.
In the rest of this section we determine the dependence of the effective 
parameters on $(H-H_c)$ due to the non--linear effects 
on scales $L\ll \xi_{v}$.  Once these dependencies are known, 
Eq. \ref{eq:edwileq} can be treated 
easily, it yields in particular in $d=3$
dimensions a logarithmic roughness of the interface for $L\gg\xi_v$. 

We start with the derivation of some scaling relations.
If the transition from the moving to the pinned state of the interface is
continuous, as we will show below, we expect a power law behavior of 
the velocity
\begin{equation}
v\sim (H-H_c)^{\tilde{\beta}}
\label{eq:velocity}
\end{equation}
and hence $\xi_{v}\sim (H-H_c)^{-\tilde{\beta}/(\tilde{z}-\tilde{\zeta})}$.
One condition for a continuous transition follows from a 
Harris--like argument: the fluctuations $\delta H(\xi_{v})$ of the 
threshold force in the correlated volume $\xi_{v}^{{d-1}+\tilde{\zeta}}$
occupied by the rough interface should be small compared to $(H-H_c)$ ~\cite
{NSTL}. Thus, 
$(H-H_c)\gg\delta H_c(\xi_{v})\sim\xi_{v}^{-({d-1}+\tilde{\zeta})/2}
\sim (H-H_c)^{\tilde{\nu}({d-1}+\tilde{\zeta})/2}$
or
\begin{equation}
\tilde{\nu}=\frac{\tilde{\beta}}{\tilde{z}-\tilde{\zeta}}\ge
\frac{2}{{d-1}+\tilde{\zeta}}
\label{eq:tildenu}
\end{equation}
where we introduced $\tilde{\nu}$ via $\xi_{v}\sim (H-H_c)^{-\tilde{\nu}}$.

For large velocities the non--linear terms of Eq. \ref{eq:mu-1}
can be treated by standard perturbation theory ~\cite{feigel}. For the renormalized inverse mobility 
we obtain e. g.
\begin{equation}
\lambda_{\rm eff}=\lambda \left( 1-c_1(\xi_{v}^{(0)}/L_c)^{5-d}+
c_2(\xi_{v}^{(0)}/L_c)^{2(5-d)}-...\right).
\label{eq:lambdaeff}
\end{equation}
Here $L_c$ is given by Eq. \ref{eq:L>Lc} with $\Delta (a)$ replaced by
$|\Delta ''(0)|a^2$ and $\xi_{v}^{(0)}=(\Gamma a/v\lambda )^{1/2}$ is
the bare dynamical correlation length.
Similar expressions can be obtained for the renormalization of $\Delta ''(0)$.
There is no renormalization of $\Gamma$, however, because
of an exact tilt symmetry~\cite{narafish}.

The perturbation theory clearly diverges for $v\rightarrow 0$.
It turns out, that a functional renormalization group calculation is 
required to find the true critical behavior since  $\Delta ''(0)$
develops a singularity under the renormalization group transformation
if $b\approx L_c$. The resulting  functional flow equation for the random--field
correlator is
\begin{equation}
\frac{{\rm d}\Delta (z)}{{\rm d}\ln{b}}=(\tilde{\epsilon}-2\tilde{\zeta})
\Delta (z)+\tilde{\zeta}z \Delta '(z) -c\frac{{\rm d}^2}{{\rm d}z^2}\left[
\frac{1}{2}\Delta^2 (z)-\Delta (z)\Delta (0)\right]
\label{eq:funcflow}
\end{equation}
where $\tilde{\epsilon}=5-d$. 
The fixed point function has a cusp--like singularity at the origin
$\Delta^{\ast}(z)=\Delta (0)+a_1|z|+a_2z^2+\ldots $
where $a_1$ and $a_2$ are of the order $\tilde{\epsilon}$.
It is important to note, that despite of the fact, that
we were able to obtained $H_c$ in Eq. \ref{eq:Hc} from a simple
scaling argument, the determination of $H_c$ from a straightforward
calculation requires the existence of a non--zero value of
$\Delta '(z\rightarrow 0^+)\sim H_c$.  The roughness exponent
$\tilde{\zeta}$ follows from the fixed point condition to be 
$\tilde{\zeta} =\tilde{\epsilon}/3$, which is identical with the equilibrium
roughness exponent. The correlator of the effective random field $h_{eff}$, 
appearing in Eq. \ref{eq:edwileq} 
is $\Delta_{eff}(z)\approx (L_c/ \xi_v)^{(\tilde{\epsilon}-2\tilde{\zeta})}
\Delta^{\ast}(z(L_c/\xi_v)^{\tilde{\zeta}})$.

The renormalization of the mobility is coupled to the renormalization of
$\Delta_{eff}(z)$ and yields $\lambda_{eff} \sim (L_c/\xi_v)^{2\tilde{\epsilon}/9}$.
Together with the scaling relations~\cite{NSTL}
\begin{equation}
\tilde{\nu}=\frac{1}{2-\tilde{\zeta}}
\label{eq:scalrel}
\end{equation}
which is valid to all orders in $\tilde{\epsilon}$~ \cite{narafish}, this gives for the 
exponents to order $\tilde{\epsilon}=5-d$
\begin{equation}
\tilde{\zeta}\approx\frac{\tilde{\epsilon}}{3}\,,\quad
\tilde{z}\approx 2-\frac{2}{9}\tilde{\epsilon}\,,\quad
\tilde{\beta}\approx 1-\frac{\tilde{\epsilon}}{9}.
\label{eq:zetazbeta}
\end{equation}
Narayan and Fisher~\cite{narafish} have claimed, that $\tilde{\zeta}$
is correct to all orders in $\tilde{\epsilon}$, but numerical 
calculations show deviations from this result in low dimensions~\cite{lesch}.
With these replacements, all terms in the
 Edwards-Wilkinson-equation Eq. \ref{eq:edwileq} scale as $H-H_c$. It is easy 
to see, that at the depinning fixed point an inertial term is irrelevant 
since it scales as $(H-H_c)^{\lambda}$ with 
$\lambda = 1+2\frac {z -1}{2-\zeta} >1$.

Thermal fluctuations will smear out the transition such that the
velocity is non--zero for all driving
forces $HM$. For $H \approx H_c$

\begin{equation}
v\approx T^{\tilde{\beta}/\rho}\psi ((H-H_c)/T^{1/\rho})
\label{eq:vTpsi}
\end{equation}
with $\psi (x)\sim x^{\tilde{\beta}}$ for $x\rightarrow\infty $ and
$\psi (x)\sim {\rm const}$ for $x\rightarrow 0$. The value of
$\rho$ is at present unknown.
For $H\ll H_c$, the velocity is exponentially small and results from a creep
motion over the barriers formed by the random field 
\begin{equation}
v\sim \exp{(-C M h^2/H T)}
\label{eq:ve}
\end{equation}
with $C=O(1)$ ~\cite{N90}.

\subsection{Critical dynamics}\label{subsec:critdyn}
Villain~\cite{VII} and D.S. Fisher~\cite{DSF86b} have 
proposed to consider the 
spin dynamics close to $T_c(h)$ as due to a kind of domain reversal, where now
domains within domains should be taken into account through the
exponent $\theta$ introduced in Eq. \ref{eq:nu} and Eq. \ref{eq:xi}.
The reversal of domain of size $\xi$ (the correlation length) is associated
to an energy barrier of order $h\xi^{\theta}$, and the Arrhenius
law gives the following expression of the relaxation time
\begin{equation}
t_{\rm rel}\approx \exp ({\rm const.}\left|\frac{T_c}{T-T_c}\right|^{\nu\theta}).
\end{equation}

This formula is in contrast with usual power laws and reminiscent of the
Vogel--Fulcher law observed in amorphous systems~\cite{joff}.
We note, that conventional perturbation theory gives
$t_{\rm rel}\approx |T-T_c|^{-\nu z}$ with dynamical critical
exponent $z=2+2\eta$ ~\cite{cardy}. Because of the exponential increase 
of the relaxation time, random field systems will rapidly fall out of
equlibrium by approaching the transition.

\subsection{Metastable domains}\label{subsec:metastabledomains}

According to our finding of the Sec. \ref{subsec:ferromphase},
the $3$--dimensional random field Ising model
should exhibit ferromagnetic order at low temperatures.
Experiments on random field systems are mainly performed with
diluted antiferromagnets in external field and
show pronounced hysteresis effects ~\cite{reviews3,reviews2,FishAha}.
In particular, no long--range order is found for $3d$ systems 
cooled from the high--temperature phase in non--zero field $H$.
Neutron scattering experiments in three dimensions yield a non-equilibrium
correlation length
\begin{equation}
R_c(t)\sim H^{-\nu_H}\cdot r(t)\quad\mbox{with $\nu_H\approx 2.1-2.2$}.
\label{eq:Rct}
\end{equation}
The absence of long--range order has been traced back in these cases to domain
wall pinning in metastable states~\cite{bruins,VI,vill}.

A short discussion of the domain relaxation ist found in~\cite{reviews1},
which gives the following prediction for the non--equilibrium correlation
length~\cite{VI}
\begin{equation}
R_c(t)\approx \frac{\sigma\Gamma}{h^2}\left(C+\frac{T}{\Gamma}{\rm ln}(t/t_0)
\right)
\end{equation}
where $C$ is a constant of order unity. This logarithmic growth has been seen 
experimentally~\cite{Feng}. In order to explain the experimental data 
of diluted antiferromagnets in an external field  at low temperatures, it is 
necessary to take into account also pinning by random bonds~\cite{vill}.

\section{Miscellaneous}

With the present review I have tried to summarize some of the more 
recent activities in the theoretical investigation of the Ising model in 
a random field. Clearly, because of limitation of space and time not all 
new and interesting developments found their place in this review. 

To mention one, Dahmen, Sethna and co-workers~\cite{Dahmen} have
studied the zero temperature random field Ising model as a model for 
noise and avalanches in hysteretic systems. Changing the external field 
$H(t)$ adiabatically from its initial state where all spins point 
downwards, at small disorder (i.e. for small $h$) the first spin to flip 
easily pushes over its neighbors, and the transition happens in one burst
 (an infinite avalanche). On the other hand, at large enough disorder 
the coupling between spins becomes negligible, and most spins flip by 
themselves, no infinite avalanche occurs.   Tuning the amount of 
disorder in the system, i.e. the random field strength $h$, one 
finds a non-equilibrium critical point ${\tilde h}_c, {\tilde H}_c$ 
where infinite avalanches disappear. 
At this point there is a universal scaling law for the magnetization 
$m=M - M_c({\tilde h}_c)$

\begin{equation}
m\approx|h-{\tilde h}_c|^{\beta} {\cal M}((H-{\tilde H}_c)/(h-{\tilde h}_c)^{\beta\delta}).
\end{equation}

The critical exponents are believed to be those obtained for the equilibrium 
random field Ising model by dimensional reduction~\cite{Dahmen}. The 
relation between the critical field $H_c$ of Section 4.1 and ${\tilde H}_c$ 
is not clear at present.

Other work not mentioned in this review considers the Ising model in a 
random field in $d=1$ dimension or on special lattices, on which exact 
solutions are possible. I also left out the random field Ising model 
in a transverse field, 
the $m \rightarrow \infty$ limit of the random field Ising model etc.. 
We refer here the reader to the earlier review 
with P. Rujan~\cite{reviews1} or to the ``Current Contents''.

\bigskip

The author of this review apologizes to all colleagues whose contribution 
has been omitted. He is conscious of its deficiencies, but nevertheless  
he hopes that it may be  of some value for a first rough 
orientation in this field.

\section*{Acknowledgments}
This research was supported by a grant from the G.I.F, the German-Israeli-Foundation 
for Scientific Research and Development. The author acknowledges his 
interaction with S. Mukherji, H. Rieger, S. Scheidl and M. Schwartz. 
In 
particular, he is very grateful to U. M\"ussel for drawing the figures.



\section*{References}
\begin{thebibliography}{99}

\bibitem{Ising}E. Ising, Z. Physik {\bf 31},253 (1925)

\bibitem{Domb}see e.g the contributions of H.N.V. Temperley and I. Syozi 
in {\em ``Phase Transitions and Critical Phenomena''}
ed. by  C. Domb and M. S. Green, Vol.{\bf 1}, 227 (1972)

\bibitem{Selke}W. Selke, in {\em ``Phase Transitions and Critical Phenomena''}
ed. by C. Domb and J. Lebowitz, Vol.{\bf 15},2 (1992)

\bibitem{Bak}P. Bak et al., \Journal{\JPC}{18}{3911}{1985}

\bibitem{Harris}A. B. Harris, \Journal{\JPC}{7}{1671}{1974}

\bibitem{Berker}K. Hui and A. N. Berker, \Journal{\PRL}{62}{2507}{1985}, 
A. N. Berker, \Journal{\PA}{194}{72}{1993}



\bibitem{BiYou}see e.g. K. Binder and A. P. Young, \Journal{\RMP}{58}{801}{1986} for a review

\bibitem{IM}Y. Imry and S.K. Ma, \Journal{\PRL}{35}{1399}{1975}

\bibitem{reviews1}T. Nattermann and J. Villain, \Journal{\PT}{11}{5}{1988}; 
T. Nattermann and P. Rujan, \Journal{\IJMP}{3}{1597}{1989}

\bibitem{reviews3}D. P. Belanger, \Journal{\PT}{11}{53}{1988}

\bibitem{reviews2}D. P. Belanger and A. P. Young, \Journal{\JMMM}{100}{272}{1991}

\bibitem{Rieger}H. Rieger, ``Monte Carlo studies of Ising spin glasses and 
random field systems'' in: `` Annual Reviews of Computational Physics, 
ed. D. Stauffer'', p.295-342, World Scientific, Singapore 1995

\bibitem{FishAha}S. Fishman and A. Aharony, \Journal{\JPC}{12}{L729}{1979}


see also D. P. Belanger, this volume

\bibitem{Vill82}J. Villain, \Journal{\JPL}{43}{808}{1982}

\bibitem{deGennes}P.G. de Gennes, J. Phys. Chem. {\bf 88}, 6469 (1984),

S. B. Dierker and P. Wiltzius, \Journal{\PRL}{58}{1865}{1987}

\bibitem{Graham}J. T. Graham et al., \Journal{\PRB}{35}{2098}{1987}

\bibitem{Fernand}J. F. Fernandez, \Journal{\EL}{5}{129}{1988}

\bibitem{Peisl}W. Fenzl and J. Peisl, \Journal{\PRL}{54}{2064}{1985}

\bibitem{Michel}C. Bostoen and K. H. Michel, \Journal{\ZPB}{71}{369}{1988},
see also T. Nattermann \Journal{\em Ferroelectrics}{104}{171}{1990} for further references

\bibitem{Gut}A. M. Gutin et al., {\em cond--mat}/9606136

\bibitem{KirkBelitz}T. R. Kirkpatrick and D. Belitz 
\Journal {\PRL} {74}{1178}{1995}

\bibitem{BE}A. Beretti, \Journal{\JSP}{38}{483}{1985}.

\bibitem{Froe}J. Fr\"ohlich, in {\em the 1984 Les Houches summer school}, 
New York, Plenum


\bibitem{Bi}K. Binder, \Journal{\ZPB}{50}{343}{1983}.

\bibitem{AizW}M. Aizenman and J. Wehr, \Journal{\PRL}{62}{2503}{1989}


\bibitem{Imb}J.Z. Imbrie, \Journal{\PRL}{53}{1747}{1984};
\Journal{\CMP}{98}{145}{1985}

\bibitem{FFS}J T. Chalker, \Journal{\JPC}{16}{6615}{1983}, 
D.S. Fisher, J. Fr\"ohlich and T. Spencer, \Journal{\JSP}{34}{863}{1984}

\bibitem{BC}J. Bricmont and A. Kupiainen, 
\Journal{\PRL}{59}{1829}{1987};
\Journal{\CMP}{}{}{1988}

\bibitem{Natter88}T. Nattermann, \Journal{\JPA}{21}{L645}{1988}

\bibitem{Natter83}T. Nattermann, \Journal{\JPC}{16}{6407}{1983}

\bibitem{AGS}A. Aharony, Y. Gefen and Y. Shapir, \Journal{\JPC}{15}{673}{1982}

\bibitem{SchVShN}M. Schwartz, J. Villain, Y. Shapir and T. Nattermann,
\Journal{\PRB}{48}{3095}{1993}

\bibitem{natter1983}T. Nattermann, \Journal{\ZPB}{54}{247}{1983}, 
{\em Phys. Stat. Sol.} (b) {\bf 132}, 125 (1985),  

\bibitem{DSF86a}D. S. Fisher, \Journal{\PRL}{56}{1964}{1986}

\bibitem{bouchaud}J. P. Bouchaud and A. Georges, \Journal{\PRL}{68}{3908}{1992}




\bibitem{grinma}V. Villain, \Journal{\JPL}{43}{L551}{1982}, 
Grinstein and S. K. Ma \Journal{\PRL}{49}{685}{1982}



\bibitem{Natter87}Y. C. Zhang, \Journal{\JPA}{19}{L941}{1986}, 
T. Nattermann, \Journal{\EL}{4}{1241}{1987}

\bibitem{Aha}A. Aharony, \Journal{\PRB}{18}{3318}{1978}

\bibitem{YoungN}A. P. Young and M. Nauenberg, \Journal{\PRL}{54}{2429}{1985}

\bibitem{Khurana}A. F. Khurana et al., \Journal{\PRL}{54}{357}{1985},
\Journal{\PRL}{55}{856}{1986}


\bibitem{RY}H. Rieger and P. Young, \Journal{\JPA}{26}{5279}{1993}

\bibitem{Gof}Gofman et al. \Journal{\PRL}{71}{1569}{1993}, \Journal {\PRB}{53}{6362}{1996}

\bibitem{Swift}M. R. Swift et al. , preprint (1996)

\bibitem{Stanley}H. E. Stanley, ``Introduction to Phase Transitions and Critical Phenomena'', Clarendon Press, Oxford 1971
\bibitem{grinstein}G. Grinstein, \Journal{\PRL}{37}{944}{1976}

\bibitem{VII}J. Villain, \Journal{\JP}{46}{1843}{1985}

\bibitem{DSF86b}D. S. Fisher, \Journal{\PRL}{56}{416}{1986}

\bibitem{SchwaSoff1}M. Schwartz and A. Soffer, \Journal{\PRL}{55}{2499}{1985}

\bibitem{SchwaSoff2}M. Schwartz and A. Soffer, \Journal{\PR}{33}{2069}{1986};
M. Schwartz et al., \Journal{\PA}{178}{6}{1991}


\bibitem{Chayes}A. B. Harris, \Journal{\JPC}{7}{1671}{1974}, J. T. 
Chayes et al. \Journal{\PRL}{57}{2999}{1986}

\bibitem{Bray}A. J. Bray and M. A. Moore \Journal{\JPC}{18}{L927}{1985}




\bibitem{DaySchY}I. Dayan, M. Schwartz and A.P. Young,
\Journal{\JPA}{26}{3093}{1993}

\bibitem{New}M. E. J. Newman et al., \Journal{\PRB}{48}{16533}{1993}

\bibitem{CaoMach}M. S. Cao and J. Machta, \Journal{\PRB}{48}{3177}{1993}

\bibitem{FalBMcK}A. Falicov, A. N. Berker and S. R. McKay,
\Journal{\PRB}{51}{8266}{1995}

\bibitem{Y}A. P. Young, \Journal{\JPC}{10}{L 257}{1977}

\bibitem{ParS}G. Parisi and N. Sourlas, \Journal{\PRL}{43}{744}{1979} 

\bibitem{LMP}D. Lancaster, E. Mariani and G. Parisi,
\Journal{\JPA}{28}{3359--73}{1995}

\bibitem{AhaIM}A. Aharony, Y. Imry and S.--K. Ma, \Journal{\PRL}{37}{1364}{1976}


\bibitem{R}H. Rieger, \Journal{\PRB}{52}{5659}{1995}


\bibitem{MY}M. Mezard and A. P. Young, \Journal{\EL}{18}{653}{1992}

\bibitem{DSF85}D. S. Fisher, \Journal{\PRB}{31}{7233}{1985}

\bibitem{MM}M. Mezard and R. Monasson, \Journal{\PRB}{50}{7199}{1994}

\bibitem{DOT}C. De Dominicis, H. Orland and T. Temesvari,
\Journal{\JdPI}{5}{987--1001}{1996}

\bibitem{dotsenko}V. Dotsenko and M. Mezard, preprint cond-mat/9611017 (1996)

\bibitem{feigel}M. V. Feigel'man, {\em Sov. Phys. JETP} {\bf 58}, 1076 (1983)

\bibitem{NSTL}T. Nattermann, S. Stepanow, L. H. Tang and H. Leschhorn
\Journal{\JdPII}{2}{1483}{1992}

\bibitem{edwards}S. F. Edwards and D. R. Wilkinson, {\em Proc. R. Soc. London},
Ser. A {\bf 381}, 17 (1982)

\bibitem{narafish}O. Narayan and D. S. Fisher, \Journal {\PRB}{48}{7030}{1993}


\bibitem{lesch}H. Leschhorn et al., {\em Annalen der Physik}, in press (1997)

\bibitem{N90}T. Nattermann, \Journal{\PRL}{64}{2454}{1990}


\bibitem{joff}J. Joffrin, in ``Ill-Condensed Matter'' (North-Holland, 1979)

\bibitem{cardy}D. Boyanowsky and J. L. Cardy, {\em Phys. Rev.} B {\bf 27}, 5557 (1983)

\bibitem{bruins}R. Bruinsma and G. Aeppli, {\em Phys. Rev. Lett.} {\bf 52}, 1543 (1983)

\bibitem{VI}



\bibitem{Feng}Q. Feng at al. \Journal{\PRB}{51}{15188}{1995}


\bibitem{vill}T. Nattermann and J. Vilfan, {\em Phys. Rev. Lett.} {\bf 61}, 223 (1988)

\bibitem{Dahmen}J. P. Sethna et al. \Journal{\PRL}{70}{3347}{1993}, 
K. Dahmen and J. P. Sethna \Journal{\PRL}{71}{3222}{1993},
\end{thebibliography}
\end{document}